# Ln$_2$(SeO$_3$)$_2$(SO$_4$)(H$_2$O)$_2$ (Ln = Sm, Dy, Yb): A Mixed-Ligand Pathway to New Lanthanide(III) Multifunctional Materials Featuring Nonlinear Optical and Magnetic Anisotropy Properties

Ebube E. Oyeka,[a] Michał J. Winiarski,[b] Hanna Świątek,[b] Wyatt Balliew,[a] Colin D. McMillen,[a] Mingli Liang,[c] Maurice Sorolla II,[d] and Thao T. Tran*[a]

[a] E.E. Oyeka, W. Balliew, Dr. C. D. McMillen, Prof. T.T. Tran
Department of Chemistry,
Clemson University,
Clemson, SC 29630, United States
E-mail: thao@clemson.edu

[b] Prof. M. J. Winiarski, H. Świątek
Faculty of Applied Physics and Mathematics and Advanced Materials Center,
Gdansk University of Technology, ul
Narutowicza 11/12, 80-233 Gdansk, Poland

[c] Dr. M. Liang
Department of Chemistry,
University of Houston,
Houston, TX 77204, United States

[d] Prof. M. Sorolla II
Department of Chemical Engineering,
University of the Philippines Diliman,
Quezon City 1101, Philippines

**Abstract:** Bottom-up assembly of optically nonlinear and magnetically anisotropic lanthanide materials involving precisely placed spin carriers and optimized metal-ligand coordination offers a potential route to developing electronic architectures for coherent radiation generation and spin-based technologies, but the chemical design historically has been extremely hard to achieve. To address this, we developed a worthwhile avenue for creating new noncentrosymmetric chiral Ln$^{3+}$ materials Ln$_2$(SeO$_3$)$_2$(SO$_4$)(H$_2$O)$_2$ (Ln = Sm, Dy, Yb) by mixed-ligand design. The materials exhibit phase-matching nonlinear optical responses, elucidating the feasibility of the heteroanionic strategy. Ln$_2$(SeO$_3$)$_2$(SO$_4$)(H$_2$O)$_2$ displays paramagnetic property with strong magnetic anisotropy facilitated by large spin-orbit coupling. This study demonstrates a new chemical pathway for creating previously unknown polar chiral magnets with multiple functionalities.

## Introduction

Lanthanide materials, exhibiting diverse physical responses when subjected to external stimuli, are at the forefront of recent technological advances in optical frequency conversion, quantum computing and spintronics.[1] Nonlinear optical lanthanide compounds are promising for improved second-harmonic generation (SHG) efficiencies and wide transparency window owing to the relatively large hyperpolarizability of Ln-based distorted polyhedra and the narrow bandwidth between the $f$ orbitals of Ln$^{3+}$ and the $s$ and $p$ orbitals of anions.[2] SHG phenomena occur in noncentrosymmetric (NCS) materials in which spatial inversion center symmetry is broken.[3] Paramagnetic Ln$^{3+}$ materials possess a large orbital magnetic moment and strong spin-orbit coupling associated with $f$ orbitals, yielding high magnetic anisotropy.[1c, 4] In addition, the "buried" nature of the 4$f$ orbitals, which is attributed to the lanthanide contraction, in tandem with their poor overlap with 5$d$ orbitals give rise to a smaller magnitude of the ligand-field-induced spitting than that of spin-orbit coupling.[5] As a result, the electronic and magnetic properties of lanthanides are mostly driven by large spin-orbit coupling, a phenomenon known to facilitate unique physical behaviors such as magnetic skyrmions, multiferroicity, quantum spin liquids, and topological states of matter.[1e, 6] Despite the impressive progress in separately investigating optical and magnetic functionalities of Ln$^{3+}$ materials, significant barriers remain as to how optical and magnetic properties can be harmonized in a single system. The targeted synergy between these multiple physical phenomena for Ln$^{3+}$ materials requires placing the Ln$^{3+}$ magnetic spins in an extended framework with the appropriate NCS lattice symmetry.[7]

The design of lanthanide compounds with a NCS crystal structure, however, constitutes a significant synthetic challenge due to the intrinsic constraints of controlling both local and extended structures simultaneously.[8] To overcome this, we combined three (SeO$_3$)$^{2-}$, (SO$_4$)$^{2-}$, and H$_2$O building units to create new NCS chiral polar lanthanide compounds.[9] In addition to framing the NCS lattice symmetry, these ligands with different electronic polarizabilities can impart favorable optical responses, offering possibilities for realizing unique multifunctional phenomena.[10]

The conventional approach to synthesizing mixed anion compounds often involves annealing a mixture of reactive oxides and halides at high temperatures or reacting those reagents in an aqueous solution of HX (X = F, Cl).[11] These synthetic methods often yield complex product mixtures, hindering chemical control of targeted materials, and thus limiting meaningful study and understanding of electron, spin, orbital, and phonon coupling.

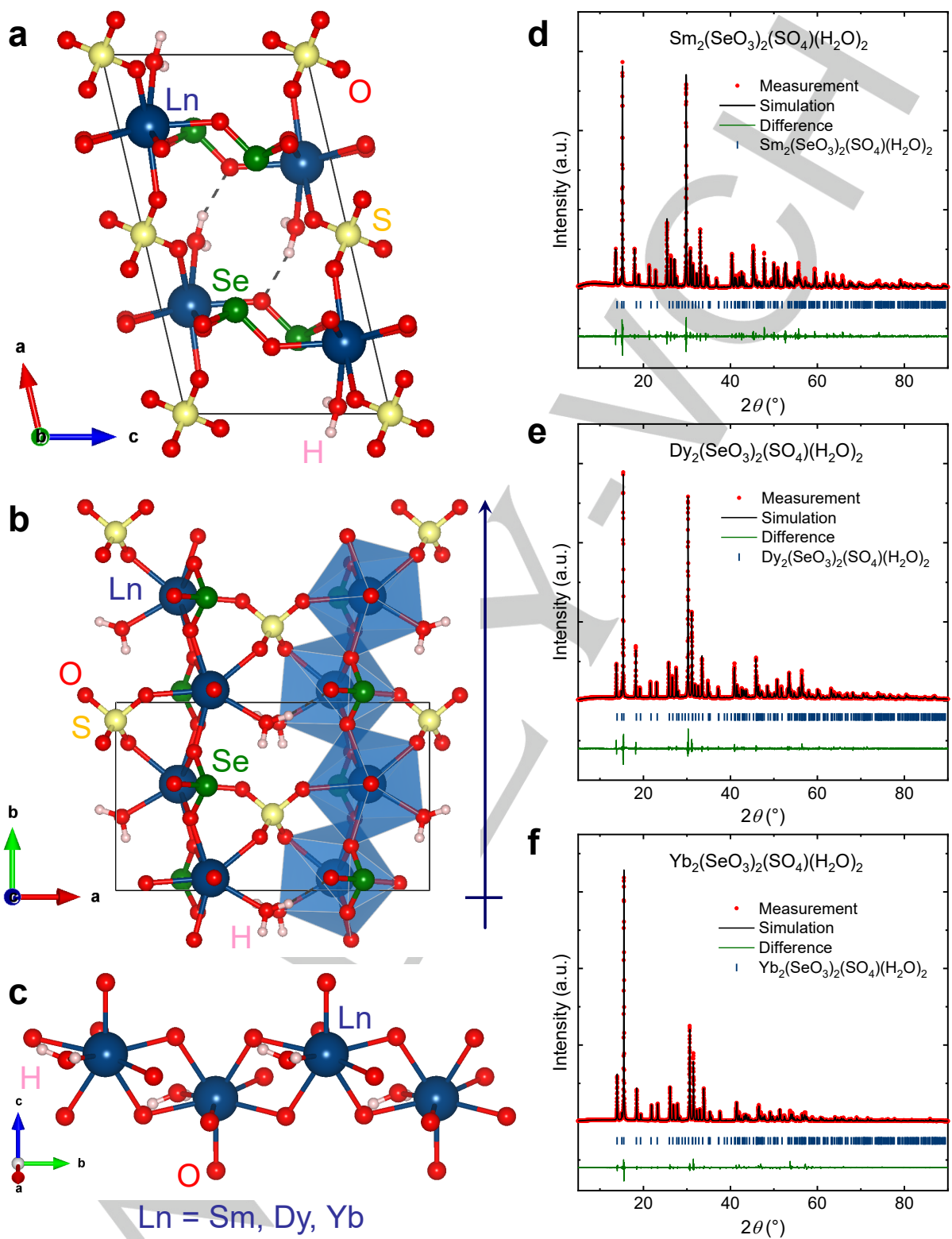

**Figure 1.** (a-b) Crystal structure of $Ln_2(SeO_3)_2(SO_4)(H_2O)_2$ consisting of chains of edge-shared $LnO_7(H_2O)$ square antiprisms, $(SeO_3)^{2-}$ trigonal pyramids, and $(SO_4)^{2-}$ tetrahedra. The $LnO_7(H_2O)$ square antiprism is aligned along the *b*-axis, resulting in the macroscopic electric polarization. (c) $LnO_7(H_2O)$ edge-sharing square antiprisms forming a 1-D zig-zag chain of $Ln^{3+}$ spins. (d-f) Rietveld refinement of powder XRD data for $Ln_2(SeO_3)_2(SO_4)(H_2O)_2$ respectively, showing the excellent agreement between the powder XRD and single crystal XRD data.

In this study, we used a thiophene acylselenourea (TAS) molecule with $C_{12}H_{10}N_2OSSe$ formula[12] as a precursor for synthesizing new chiral lanthanide $Ln_2(SeO_3)_2(SO_4)(H_2O)_2$ (Ln = Sm, Dy, Yb) materials. The combination of the selenourea moiety and thiophene ring in the precursor enables the formation of $Ln_2(SeO_3)_2(SO_4)(H_2O)_2$ using TAS as a single source for Se and S. This synthetic versatility places $Ln^{3+}$ magnetic cations in the NCS polar chiral lattice formed by three $(SeO_3)^{2-}$, $(SO_4)^{2-}$, and $H_2O$ building units, providing a chemically feasible pathway for integrating spins into desired structures. We investigated the role of electron, spin, orbital, and phonon components in $Ln_2(SeO_3)_2(SO_4)(H_2O)_2$ (Ln = Sm, Dy, Yb) and their contributions to optical and magnetic responses by performing spectroscopy, phase-matching SHG, temperature-dependent and field-dependent magnetization measurements. Furthermore, we supplement these experiments with heat capacity measurements to interrogate the thermodynamic signatures of the materials. For completion, we seek to understand the nature of orbital overlapping that gives rise to the multifunctional behaviors of $Ln_2(SeO_3)_2(SO_4)(H_2O)_2$ by performing density functional theory (DFT) calculation. This work demonstrates a step forward in the direction of atomically precise engineering of electron, spin, orbital, and phonon dynamics in $Ln^{3+}$-based NCS polar chiral architectures.

## Results and Discussion

The crystal structures of $Sm_2(SeO_3)_2(SO_4)(H_2O)_2$ and $Dy_2(SeO_3)_2(SO_4)(H_2O)_2$ were determined by single-crystal XRD (Figure 1a-c, Table S1). Rietveld refinements from room temperature powder XRD measurements were consistent with the single crystal structures and supported the phase purity of the reaction product (Figure 1d-f, Table S2-S11). $Ln_2(SeO_3)_2(SO_4)(H_2O)_2$ (Ln = Sm, Dy, Yb) materials crystallize in the NCS chiral, polar, monoclinic $C2$ space group with similar unit cell parameters, making them isostructural and isomorphic. $Ln_2(SeO_3)_2(SO_4)(H_2O)_2$ adopts the same space group as the selenite-selenate $Er(SeO_3)(SeO_4)_{0.5} \cdot H_2O$ compound.[13] The Se atom in $Ln_2(SeO_3)_2(SO_4)(H_2O)_2$ is fully ordered, whereas that in $Er(SeO_3)(SeO_4)_{0.5} \cdot H_2O$ is disordered.[13] The crystal structure of $Ln_2(SeO_3)_2(SO_4)(H_2O)_2$ features a 3D network of $(SeO_3)^{2-}$ trigonal pyramids, $(SO_4)^{2-}$ distorted tetrahedra, and $LnO_7(H_2O)$ square antiprisms (Figure 1b). Each $Ln^{3+}$ cation is coordinated to seven oxygen atoms from the selenite and sulfate groups and one from the $H_2O$ molecule, forming an eight-coordinate geometry that can be described as a distorted square antiprism. $LnO_7(H_2O)$ square antiprisms are connected in an edge-shared fashion along the $b$-axis. The $Ln^{3+}$ cations form a one-dimensional zig-zag chain sub-lattice (Figure 1c) with the Ln-Ln distances of 4.0292(4) Å, 3.9685(8) Å, and 3.8508(10) Å for Sm, Dy, and Yb materials, respectively. The top and bottom basal planes of the $LnO_7(H_2O)$ square antiprism comprise $LnO_4$ and $LnO_3(H_2O)$. The top plane is rotated by ~31° from the plane below. This value deviates from the distortion angle of 45° for an ideal square antiprism.[14] The Ln-O bond distances range from 2.349(6) – 2.478(9) Å, 2.301(6) – 2.441(12) Å, and 2.298(6) – 2.474(11) Å in Sm, Dy, and Yb compounds, respectively.

$Ln_2(SeO_3)_2(SO_4)(H_2O)_2$ structure has three building units, i.e., $(SeO_3)^{2-}$, $(SO_4)^{2-}$, and $H_2O$, through which neighboring lanthanide oxide chains are connected. $(SeO_3)^{2-}$ trigonal pyramids feature Se in +4 oxidation state with stereoactive lone-pair electrons. Each Se atom is bonded to two bridging oxygen (μ-O) and one terminal oxygen in a distorted trigonal pyramidal geometry, with Se-O bond distances ranging from 1.580(12) Å – 1.715(13) Å. The (μ-O)-Se-(μ-O) bond angle (92.8(4)° – 96.3(7)°) is smaller than the other O-Se-O bond angles (100.2(6)° – 105.5(9)°), leading to the reduction in the local symmetry of $(SeO_3)^{2-}$ from $C_{3v}$ to pseudo-$C_s$ point group. These $(SeO_3)^{2-}$ groups connect neighboring chains along the $c$-axis. The $(SO_4)^{2-}$ anion contains S with the formal oxidation state of +6, bonded to four oxygen atoms in a distorted tetrahedral geometry (pseudo-$T_d$ point group). S-O bond distances range from 1.438(14) Å – 1.503(11) Å. These connect neighboring chains along the $a$-axis. $H_2O$ ligands bond to the $Ln^{3+}$ cation through its O atom, and form Se-O---H hydrogen bonds with the O atoms of $(SeO_3)^{2-}$ unit. H-O bond length and H-O-H bond angle were fixed to 0.85(3) Å and 108(7)° respectively during the least-squares refinements, and the local symmetry of $H_2O$ ligand is $C_{2V}$. The macroscopic electric polarization of $Ln_2(SeO_3)_2(SO_4)(H_2O)_2$ results from the alignment of the polar $LnO_7(H_2O)$ square antiprism along the $b$-axis (Figure 1b). The local electric dipoles of the $(SeO_3)^{2-}$ trigonal pyramids do not add to the overall polarization because they arrange in the opposite directions. The symmetry and local structure of the $(SeO_3)^{2-}$, $(SO_4)^{2-}$, and $H_2O$ linkers were confirmed by the infrared spectra of the Ln materials (Figure S4). The UV-Vis spectra of the lanthanide compounds proved the characteristics of the localized $4f$ orbitals and strong spin–orbit (LS) coupling of the $Ln^{3+}$ ions, and negligible ligand-field-induced splitting (Figure S5).

Motivated by the NCS chiral polar structure of $Ln_2(SeO_3)_2(SO_4)(H_2O)_2$, we examined the second-order nonlinear optical properties. To estimate more accurately SHG intensity and evaluate phase-matching behavior for the materials, SHG efficiency measurements at 1064 nm as a function of particle size were performed. These measurements yielded a SHG efficiency of nearly 0.05, 0.04, and 0.12 × $KH_2PO_4$ (KDP, the SHG phase-matching standard material at 1064 nm) for the Sm, Dy, and Yb material, respectively, in the 45 – 63 μm particle size range (Figure 2). The low SHG intensity of the materials can be attributed to three possible factors: the arrangement of the anion groups, the narrow bandwidth between the $f$ states of $Ln^{3+}$ and the $s$ and $p$ states of the ligands, and unfavorable electronic transitions of the partially filled $f$ orbitals of the $Ln^{3+}$ cations.

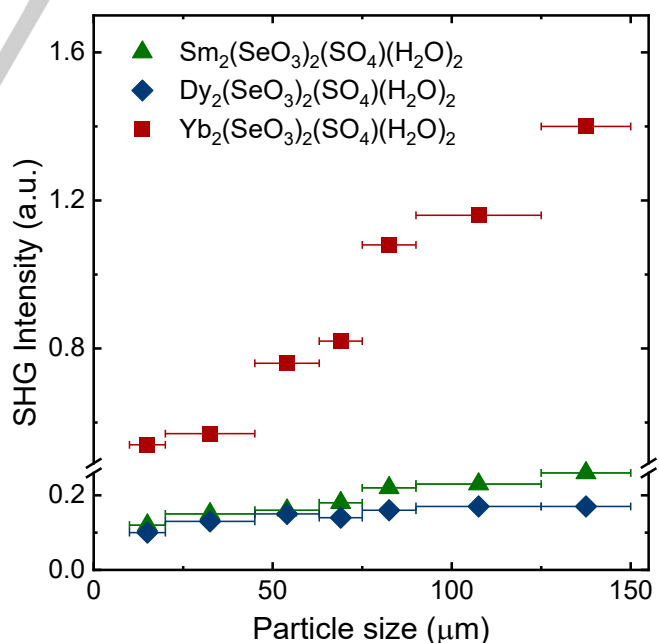

**Figure 2.** SHG intensity of $Ln_2(SeO_3)_2(SO_4)(H_2O)_2$ (Ln = Sm, Dy, Yb) as a function of particle size.

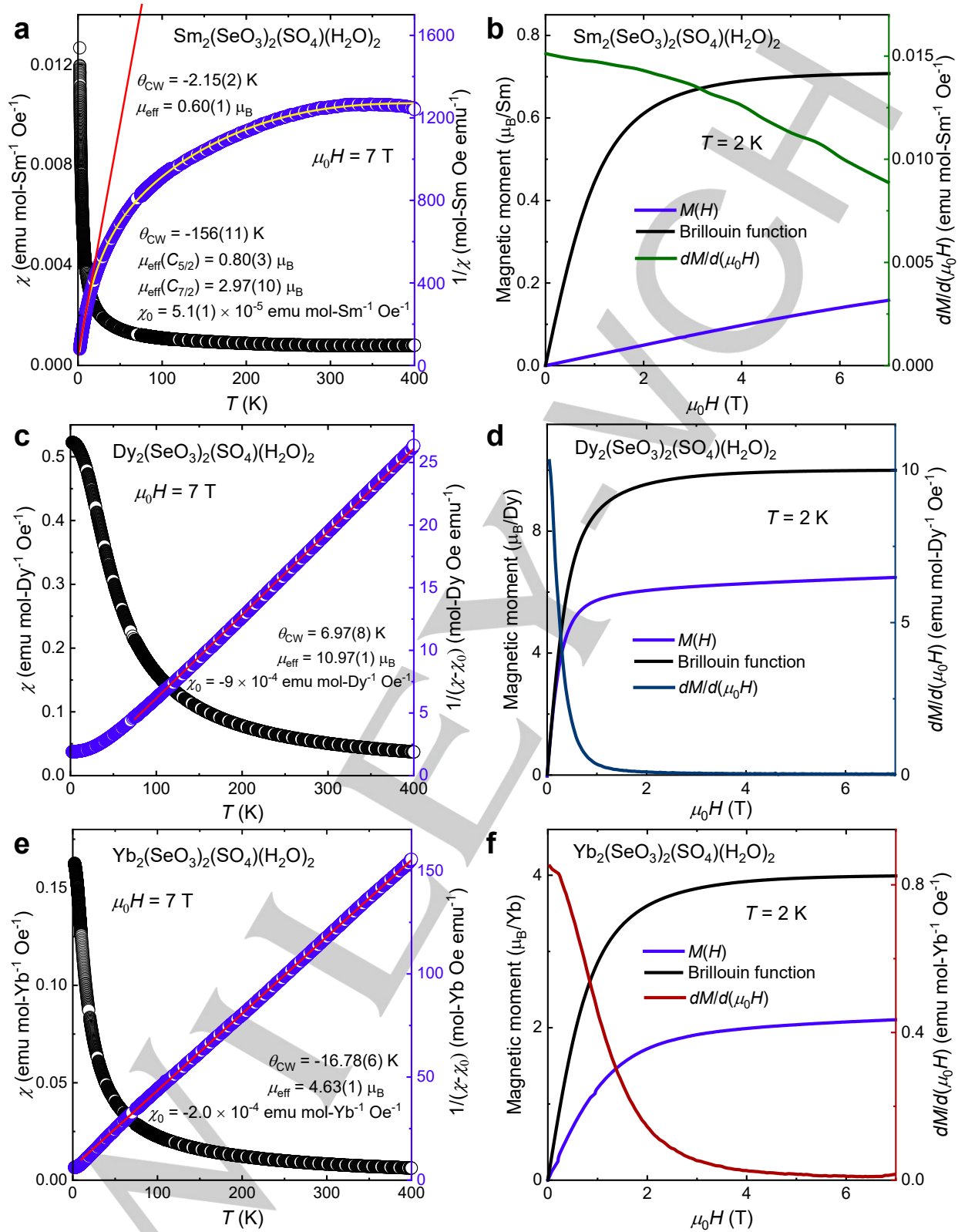

**Figure 3.** (a,c,e) $Ln_2(SeO_3)_2(SO_4)(H_2O)_2$ (Ln = Sm, Dy, and Yb, respectively) dc magnetic susceptibility ($\chi$) from $T$ = 2 K – 400 K at $\mu_0H$ = 7 T (black); Curie-Weiss fit of $1/(\chi-\chi_0)$ vs $T$ (red line); analysis of $1/\chi$ vs $T$ of the Sm material using the modified Curie-Weiss equation (green line). (b,d,f) $M(H)$ data in $\mu_B$ at $T$ = 2 K for the Sm, Dy, and Yb system, respectively; Brillouin function for $J$ = 5/2, 15/2, and 7/2 and $T$ = 2 K; $dM/dH$ curve showing no critical field.

The Yb$^{3+}$ ($f^{13}$) material possesses the strongest SHG efficiency among the Ln$^{3+}$ compounds, 2.4 × Sm$^{3+}$ ($f^5$) and 3 × Dy$^{3+}$ ($f^9$). This observation can be linked to the less likelihood of having unfavorable electronic transitions for Yb$^{3+}$ ($f^{13}$) between the ground state ($^2F_{7/2}$) and virtual excited states, in comparison to the $f^5$ and $f^9$ electronic structures of the other two members of this family. As the particle size increases, the SHG intensity of Ln$_2$(SeO$_3$)$_2$(SO$_4$)(H$_2$O)$_2$ increases and reaches a plateau (Figure 2). This behavior indicates that the Ln$^{3+}$ materials exhibit type 1 phase-matching behavior and fall into the class A category of SHG materials.[15] The phase-matching property of the Ln$^{3+}$ materials reveals that the fundamental and harmonic waves travel at the same propagation speed. This result proved that the combination of the electronic structure and lattice symmetry of the new NCS Ln$_2$(SeO$_3$)$_2$(SO$_4$)(H$_2$O)$_2$ system works in favor of harmonizing the wave-vectors and the phases of fundamental and second-harmonic lights, a protocol for effectively generating coherent light through the SHG process.

To evaluate the spin and orbital magnetic contribution of the Ln$^{3+}$ cation in Ln$_2$(SeO$_3$)$_2$(SO$_4$)(H$_2$O)$_2$, we performed magnetization measurements. The $\chi(T)$ curve of Sm$_2$(SeO$_3$)$_2$(SO$_4$)(H$_2$O)$_2$ features paramagnetic behavior down to $T$ = 2 K (Figure 3a). In contrast to most other lanthanide (III) cations, the energy separation between the $^6H_{5/2}$ ground state and the $^6H_{7/2}$ first excited state of Sm$^{3+}$ is comparable with $kT$, resulting in the $^6H_{7/2}$ state being thermally populated and mixing significantly with the ground state.[16] Thus, the inverse magnetic susceptibility $1/\chi$ versus temperature of the Sm material at high temperature 20 K ≤ $T$ ≤ 400 K cannot be explained by the conventional Curie-Weiss equation owing to non-negligible magnetic contribution from the $^6H_{7/2}$ excited state.[1e] Similar behavior has also been observed in other Sm$^{3+}$ compounds such as SmPd$_2$Al$_3$, SmB$_4$, and KBaSm(BO$_3$)$_2$.[1e, 6b, 17] The magnetic susceptibility of Sm$_2$(SeO$_3$)$_2$(SO$_4$)(H$_2$O)$_2$ can be estimated by the modified Curie-Weiss equation (equation 1):[1e]

$$\chi(Sm^{3+}) = \chi_0 + [C_{5/2}/(T-\theta_{CW})][1 - e^{-\Delta/T}] + (C_{7/2}/T)e^{-\Delta/T} \quad (1)$$

where $C_{5/2}$ and $C_{7/2}$ are the Curie constants for the $^6H_{5/2}$ and $^6H_{7/2}$ states, respectively, $\chi_0$ is the temperature-independent contribution to the susceptibility, and $\theta_{CW}$ is the Curie-Weiss temperature. From the fit of $1/\chi$ versus $T$ curve (Figure 3a), the Curie constants were extracted to be $C_{5/2}$ = 0.080(7) emu K mol-Sm$^{-1}$ and $C_{7/2}$ = 1.10(8) emu K mol-Sm$^{-1}$, and the 'Curie-Weiss temperature' $\theta$ = -156(11) K is indicative of the crystal-field splitting energy rather than the exchange energy.[18] The obtained effective magnetic moment for the $^6H_{5/2}$ and $^6H_{7/2}$ states are 0.80(3) $\mu_B$ and 2.97(10) $\mu_B$, respectively. These values are consistent with the effective moment for an isolated Sm$^{3+}$ cation ($\mu_{eff}$($^6H_{5/2}$) = 0.85 $\mu_B$; $\mu_{eff}$($^6H_{7/2}$) = 2.54 $\mu_B$).[1e] At low temperature, the statistical occupation of the $^6H_{7/2}$ excited state is less pronounced, thus the magnetic susceptibility of the Sm material can be analyzed using the Curie-Weiss law (equation 2).[19]

$$\chi(T) = \frac{C}{T - \theta_{CW}} + \chi_0 \quad (2)$$

where $C$ is the Curie constant, $\theta_{CW}$ is the Curie-Weiss temperature, and $\chi_0$ is the temperature-independent contribution to the susceptibility, including the small diamagnetic signals of the electron core and the sample holder.[20]

A linear correlation between inverse susceptibility $1/(\chi-\chi_0)$ and temperature indicates the Curie-Weiss character. The effective magnetic moment $\mu_{eff}$ per Ln$^{3+}$ cation was estimated using the relation (equation 3):

$$\mu_{eff} = \sqrt{\left(\frac{3k_B}{N_A\mu_B^2}\right)C} \quad (3)$$

where $N_A$ is the Avogadro number, $k_B$ is the Boltzmann constant and $\mu_B$ is the Bohr magneton. A fit of the low-temperature part of the inverse susceptibility (4 K ≤ $T$ ≤ 10 K) yields an effective moment $\mu_{eff}$ = 0.60(1) $\mu_B$ and a Weiss temperature $\theta_{CW}$ = -2.15(2) K, consistent with the lack of magnetic ordering observed down to $T$ = 1.9 K.[1e, 6b, 18, 21] Although the $\mu_{eff}$ effective moment is smaller than the expected value of 0.85 $\mu_B$ for Sm$^{3+}$ free ion, it is similar to the magnetic moment obtained for other relevant Sm$^{3+}$ materials.[22] The relatively low effective moments observed in these samarium compounds are likely connected to the strong crystal field effect of the $^6H_{5/2}$ multiplet. To understand how magnetization evolves as a function of magnetic field, we measured field-dependent magnetization from $\mu_0H$ = 0 T – 7 T at $T$ = 2 K (Figure 3b). The Brillouin function illustrates magnetization of an ideal paramagnet where the spin interaction is negligible.[8b, 23] The $M(H)$ curve of the Sm compound only approaches ~20% of the saturation moment estimated by the Brillouin function, suggesting significant orbital contribution and high magnetic anisotropy.[24] The $dM/dH$ curve exhibits no field-induced transition for Sm$_2$(SeO$_3$)$_2$(SO$_4$)(H$_2$O)$_2$ (Figure 3b).

The magnetic susceptibility of the Dy material at $\mu_0H$ = 7 T is presented in Figure 3c. The data of the paramagnetic region at 72 K ≤ $T$ ≤ 400 K were analyzed using the Curie-Weiss equation (equation 2). The effective magnetic moment obtained for Dy$^{3+}$ cation is 10.97(1) $\mu_B$/Dy$^{3+}$, which is close to the $g[J(J+1)]^{1/2}$ = 10.65 $\mu_B$ expected for a $J$ = 15/2 moment.[6b] This indicates the strong interaction between the quantum numbers $S$ and $L$, consistent with the results from the electronic spectra of the Dy material. The Curie-Weiss temperature $\theta_{CW}$ of Dy$_2$(SeO$_3$)$_2$(SO$_4$)(H$_2$O)$_2$ was calculated to be 6.97(8) K. The relatively small positive value of $\theta_{CW}$ indicates weak ferromagnetic spin-spin interactions in the Dy material. The $M(H)$ curve (Figure 3b) at $T$ = 2 K shows nonlinear behavior with an onset starting M ≈5.7 $\mu_B$/Dy at $\mu_0H$ ≈ 1 T, and reaches the value of 6.5 $\mu_B$ at $\mu_0H$ = 7 T. The Brillouin function, which describes non-interacting paramagnetic spins, were calculated for Dy$^{3+}$ ($J$ = 15/2) at $T$ = 2 K (Figure 3d). Although the $M(H)$ curve of Dy$_2$(SeO$_3$)$_2$(SO$_4$)(H$_2$O)$_2$ shows a similar trend as the Brillouin function, its departure from the ideal paramagnet model implies non-negligible spin-spin interactions and high magnetic anisotropy present in the material. Furthermore, the magnetic saturation of the Dy compound approaches about 60% of the expected maximum value due to magnetic anisotropy. Derivatives of the $M(H)$ curve display no anomaly feature (Figure 3d), consistent with the absence of a finite-field phase transition at T = 2 K. In addition, $dM/dH$ declines rapidly and comes down to zero at $\mu_0H$ ≥ 2 T, which is in congruence with the magnetic saturation observed in the $M(H)$ curve.

The magnetic susceptibility data of the Yb compound (at $\mu_0H$ = 7 T) follow the Curie-Weiss law over a wide temperature range 10 K ≤ T ≤ 400 K (Figure 3e).[6b] The effective magnetic moment obtained for the Yb$^{3+}$ cation is $\mu_{eff}$ = 4.63(1) $\mu_B$/Yb$^{3+}$, which is approximately the moment for a $J$ = 7/2 system ($g[J(J+1)]^{1/2}$ = 4.54 $\mu_B$). This suggests significant $L$-$S$ coupling, in concert with the conclusion deduced from the electronic spectra of the Yb compound.

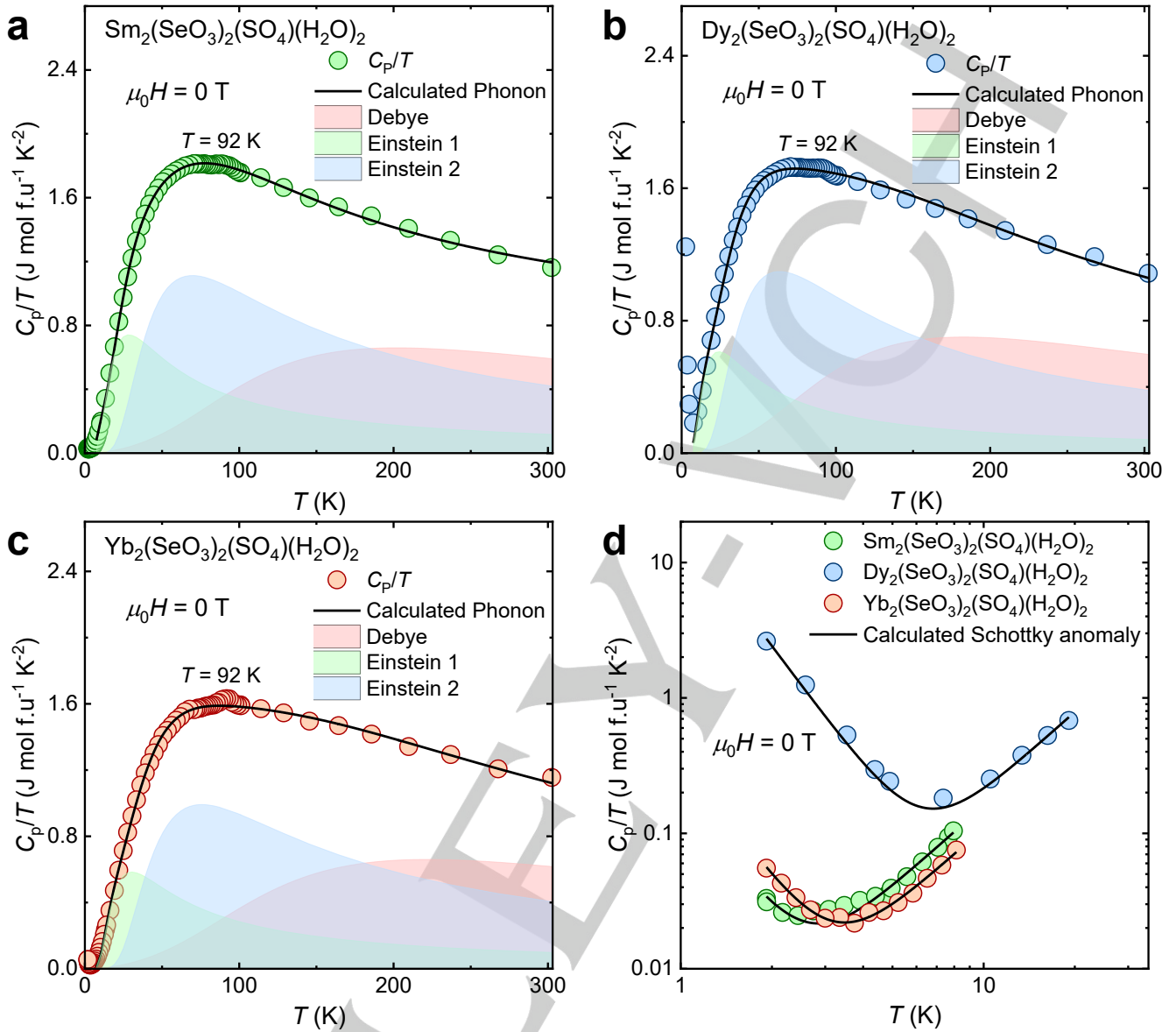

**Figure 4.** (a,b,c) Molar heat capacity over temperature ($C_P/T$) versus temperature ($T$) and phonon estimation for $Ln_2(SeO_3)_2(SO_4)(H_2O)_2$ (Ln = Sm, Dy, and Yb, respectively). The calculated phonons were best described using a combination of one Debye and two Einstein modes (equation 6). The resulting parameters are presented in Table 1. (d) Analysis of Schottky anomaly for the Sm, Dy, and Yb material at $T \leq 11$ K.

No indication of a transition to a long-range ordered magnetic state was observed down to $T$ = 2 K and the Curie-Weiss temperature $\theta_{CW}$ = -16.78(6) K, implying antiferromagnetic interaction between the $Yb^{3+}$ spins, yields a frustration parameter of approximately 8.4.[6b] This result indicates that the 1D chain of Yb–Yb with relatively short distance of 3.8508(10) Å combines with large SOC in the $Yb_2(SeO_3)_2(SO_4)(H_2O)_2$ chiral magnet to potentially give rise to unusual magnetic states. Similar phenomena of strongly correlated quantum states such as quantum spin liquid and spin ice have been realized in geometrically frustrated lattices of $Yb^{3+}$ $S$ = ½ spins in $YbMgGaO_4$, $KYb_2F_5SO_4$, $LiYbSe_2$, $NaYbL_2$ (L = O, S), and $YbBO_3$[7a, 25]. While $Yb_2(SeO_3)_2(SO_4)(H_2O)_2$ can make an appealing system for exploring highly entangled quantum states, further studies such as dc and ac magnetization at $T$ < 2K using dilution refrigerator and neutron scattering experiments are needed to gain deeper insight into the magnetic ground state of this material.

The $M(H)$ curve measured at $T$ = 2 K exhibits nonlinear behavior and approaches magnetic saturation M ≈ 2.0 $\mu_B$/Yb at $\mu_0H$ ≈ 7 T (Figure 3f). The Brillouin function, modelling the magnetic moment of $Yb^{3+}$ non-interacting spins at $T$ = 2 K, indicates that the $M(H)$ curve should ideally saturate at around 4 $\mu_B$. The magnetic saturation of the Yb compound is observed at approximately 50% of the expected value, confirming significant orbital contribution to the magnetic moment. While the trend of the $M(H)$ curve of the Yb material resembles that of the Brillouin function, the deviation from the ideal paramagnetic model infers antiferromagnetic spin-spin interaction in the system.[6b, 25c]

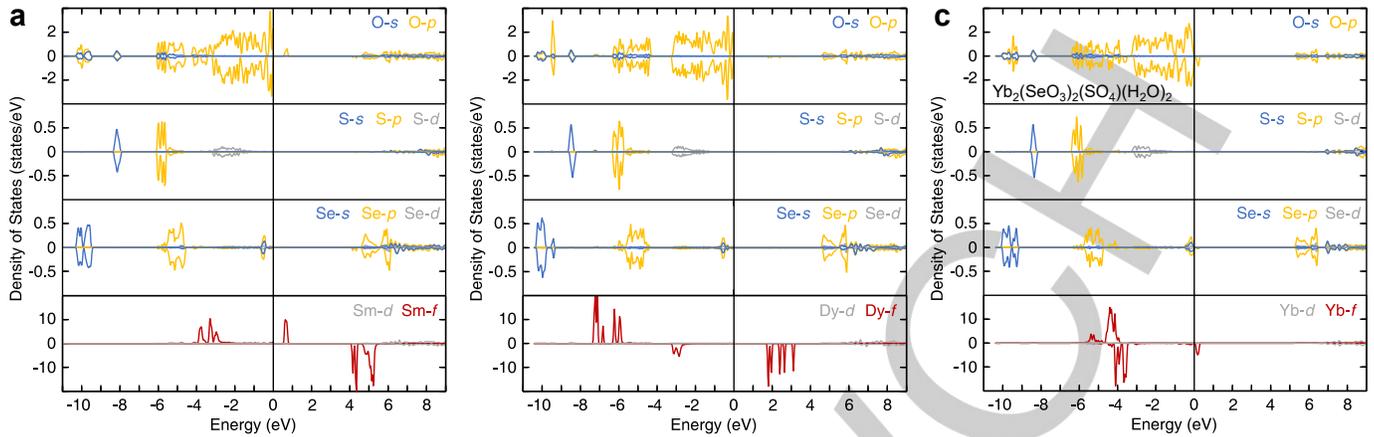

**Figure 5.** (a-c) Spin-polarized density of states (DOS) of $Ln_2(SeO_3)_2(SO_4)(H_2O)_2$ (Ln = Sm, Dy, and Yb, respectively) showing orbital overlap.

The $dM/dH$ curve decreases to approximate zero at higher magnetic fields (Figure 3f), in an excellent agreement with the magnetic saturation seen in the $M(H)$ curve.

To study electronic and phononic contributions for $Ln_2(SeO_3)_2(SO_4)(H_2O)_2$, we performed heat capacity measurements at $\mu_0H$ = 0 T over the temperature range of 2 K ≤ $T$ ≤ 300 K. The $C_P/T$ versus $T$ curves are presented in Figure 4. A small kink at $T$ ≈ 92 K is observed in all three materials, suggesting a structural phase transition. This may be attributable to disorder and order of $H_2O$ in the structure, not affecting the magnetic behavior. The crystal structure obtained from the single-crystal XRD data collected at $T$ = 100 K (Table S3 in Supporting Information) is identical with that deduced from the data measured at $T$ = 300 K. Probing the structural phase transition in $Ln_2(SeO_3)_2(SO_4)(H_2O)_2$ necessitates a diffraction experiment at $T$ < 90 K, which will be performed in a separate study. There is no $T$-linear contribution to the specific heat data, confirming no electronic density around the Fermi level. This is consistent with the insulating nature of the $Ln^{3+}$ materials. As aforementioned, no magnetic transition was observed in the magnetization data of the $Ln^{3+}$ compounds. Thus, the total heat capacity of $Ln_2(SeO_3)_2(SO_4)(H_2O)_2$ is attributed to phonon contributions. The specific heat data were best modeled using a combination of one Debye and two Einstein modes (Figure 4a-c) (Equation 4). Equations 5 and 6 describe the Debye and Einstein models, respectively.[26]

$$C_P(T) = C_D(\theta_D, s_D, T) + C_{E1}(\theta_{E1}, s_{E1}, T) + C_{E2}(\theta_{E2}, s_{E2}, T) \quad (4)$$

$$C_D(\theta_D, T) = 9s_D R \left(\frac{T}{\theta_D}\right)^3 \int_0^{\theta_D/T} \frac{(\theta_D/T)^4 e^{\theta_D/T}}{[e^{(\theta_D/T)}-1]^2} d\frac{\theta_D}{T} \quad (5)$$

$$C_E(\theta_E, T) = 3s_E R \left(\frac{\theta_E}{T}\right)^2 \frac{e^{\theta_E/T}}{[e^{(\theta_E/T)}-1]^2} \quad (6)$$

where $\theta_D$ is the Debye temperature; $\theta_{E1}$ and $\theta_{E2}$ are the Einstein temperatures; $s_D$, $s_{E1}$, and $s_{E2}$ are the oscillator strengths; and $R$ is the molar gas constant. The resulting parameters are summarized in Table S12 of the Supporting Information. The sum of the oscillators is 16.06(44), 14.75(28), and 16.60(43) for $Ln_2(SeO_3)_2(SO_4)(H_2O)_2$ (L = Sm, Dy, and Yb, respectively). These values are smaller than the expected value of 21, which is the number of atoms per formula unit. The underestimated results are likely owing to the coupling of the lattice excitations, which can be driven by strong covalent bonding characters of the $SeO_3$, $SO_4$ and $H_2O$ groups.

The upturn in the specific heat data of $Ln_2(SeO_3)_2(SO_4)(H_2O)_2$ at $T$ < 4 K indicates the onset of a Schottky anomaly, arising from the paramagnetic $Ln^{3+}$ spins. The $C_p$ data at $T$ ≤ 11 K were analyzed using equation 7 (Figure 4d, Table S13).[26a, 27]

$$C_P/T = \beta_3 T^2 + AT^{-3} \quad (7)$$

where $\beta_3$ is the thermal expansion coefficient which corresponds to the lattice vibration, and $A$ is the Schottky parameter which is related to the splitting energy between the nondegenerate energy levels. No electronic contribution ($\gamma$) was included in the expression attributed to the insulating behavior of $Ln_2(SeO_3)_2(SO_4)(H_2O)_2$. The calculated $\beta_3$ lattice vibration values of the three compounds are of the same magnitude because they are isostructural (Table S13). It is worth noting that the determination of the Schottky parameter is not perfect since only the beginning of the anomaly was observed down to $T$ = 2 K. However, taking the Schottky term into account suggests that the upturn in the specific heat data is not indicative of an onset of a magnetic phase transition, but rather the Schottky effect expected from the $Ln^{3+}$ paramagnetic systems.

To tie the aforementioned physical properties and crystal structure of $Ln_2(SeO_3)_2(SO_4)(H_2O)_2$ to chemical bonding and electronic structure, density functional theory (DFT) computations were performed using the Wien2K code (Figure 5).[28] The spins of Ln-$f$ states are polarized, and then polarize the O-$s$, O-$p$, S-$s$, S-$p$, Se-$s$, and Se-$p$ electrons, giving rise to magnetic interactions. The valence band maximum is mainly composed of Se-$s$, Se-$p$, and O-$p$ states, while the conduction band minimum is mostly characterized by the Ln-$f$ states with small contribution of O-$p$ orbitals. Despite the localized characteristics of the Ln $4f$ orbitals, these $f$ states hybridize with O-$p$ states. This overlap aids the synchronization of optically nonlinear and magnetically anisotropic phenomena in the Sm, Dy, and Yb mixed-anion NCS chiral magnets.

**Conclusion**

Engineering targeted NCS frameworks of paramagnetic $Ln^{3+}$ systems is critical for harmonizing optical and magnetic responses, a prerequisite for powerful new capabilities for multifunctional technologies, but this is not a trivial task. Mixed $SeO_3^{2-}$, $SO_4^{2-}$ and $H_2O$ ligands were chosen based on their different electronic polarizabilities and local symmetries to template a resulting NCS structure, enabling the creation of new NCS chiral polar magnets $Ln_2(SeO_3)_2(SO_4)(H_2O)_2$ (Ln = Sm, Dy, Yb). The materials are capable of generating coherent light with double energy of the incident radiation. Their SHG phase-matching behavior proved the viability of the heteroanionic design consideration for breaking spatial inversion symmetry in the structure. The magnetic properties of the compounds are driven by strong spin-orbit coupling and large magnetic anisotropy, a protocol for new spin physics. The $Ln^{3+}$ materials feature no magnetic phase transition down to $T$ = 2 K and only a Schottky anomaly attributed to the $Ln^{3+}$ paramagnetic spins. By creating the NCS framework, we demonstrated the nature of primarily ligand-based phonons implicated in the collective lattice excitations owing to strong covalent bonding characters of the $SeO_3^{2-}$, $SO_4^{2-}$, and $H_2O$ linkers. The results elucidate the mixed ligand design as a worthwhile avenue for integrating spins in tailored frameworks and modifying coupled degrees of freedom in these NCS chiral polar magnets. In spite of the 'buried' characteristics of the 4$f$ orbitals, these $f$ states overlap with the O-$p$ states, facilitating the unique harmonization of SHG phase-matching behavior and magnetic anisotropy in the single system. This work offers a crucial roadmap for electron, spin, orbital, and phonon engineering in multifunctional materials to simultaneously tune optical and magnetic properties at the atomic level.

**Acknowledgements**

This work was supported by Clemson University, College of Science, Department of Chemistry. The National Science Center (Poland) supported research at Gdansk University of Technology under the SONATA-15 grant (no. 2019/35/D/ST5/03769). MJW gratefully acknowledges the Ministry of Science and Higher Education scholarship for young scientists. We thank Dr. R. Sachdeva for his assistance in the FTIR measurements. We greatly appreciate the Halasyamani group for the SHG measurements.

**Keywords:** lanthanide • mixed anions • magnetic anisotropy • nonlinear optical • asymmetric ligands

charge by the joint Cambridge Crystallographic Data Centre and Fachinformationszentrum Karlsruhe Access Structures service.

**Entry for the Table of Contents**

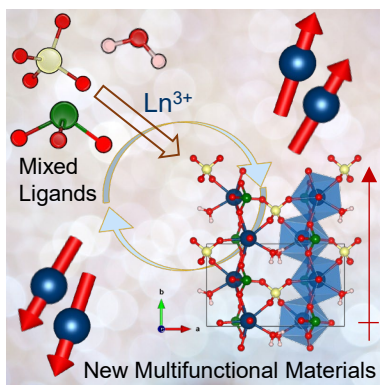

Our work provides a mixed-ligand gateway to creating new lanthanide materials having multiple functionalities, which often do not coexist. $Ln_2(SeO_3)_2(SO_4)(H_2O)_2$ materials are capable of generating coherent light through the second-harmonic generation and exhibiting magnetic anisotropy facilitated by large spin-orbit coupling. This result shares new ideas for harmonizing optical and magnetic properties in single systems at the atomic level.

@MatChemWhisper @EbubeOyeka